# Intrinsic Electronic Defect States of Anatase using Density Functional Theory

Abhishek Raghav,[1] Adie Tri Hanindriyo,[2] Keishu Utimula,[2] Mohaddeseh Abbasnejad,[3] Ryo Maezono,[4] and Emila Panda[1,a]

[1]*Department of Materials Science and Engineering, Indian Institute of Technology, Gandhinagar, Palaj 382355, Gujarat, India*

[2]*School of Materials Science, Japan Advanced Institute of Science and Technology (JAIST), 1-1 Asahidai, Nomi, Ishikawa, 923-1292, Japan*

[3]*Faculty of Physics, Shahid Bahonar University of Kerman, Kerman, Iran*

[4]*School of Information Science, JAIST, 1-1 Asahidai, Nomi, Ishikawa, 923-1292, Japan*
 *Computational Engineering Applications Unit, RIKEN, 2-1 Hirosawa, Wako, Saitama 351-0198, Japan*

## ABSTRACT

In this work an overall electronic structure including the position and formation energies of various intrinsic defects are computed for anatase using Density Functional Theory aided by Hubbard correction (DFT+U). The intrinsic point defects considered here are, oxygen vacancy ($V_O$), oxygen interstitial ($O_i$), titanium vacancy ($V_{Ti}$) and titanium interstitial ($Ti_i$). Out of all the intrinsic defects considered here, $V_{Ti}$ and $Ti_i$ are found to be most stable under equilibrium condition. Whereas, conduction band in anatase is consisted of mainly Ti 3d with a minor component of O 2p states, valence band is found to be mainly composed of O 2p with a minor contribution from Ti 3d states. $V_O$ and $Ti_i$ are found to form localized states in the band gap. Moreover, anisotropy in the effective mass is seen. Finally, an alignment of band diagrams for all the intrinsic defect states is performed using vacuum potential from slab-supercell calculation as reference. This first principle study would help in the understanding of defect-induced insulating to conducting transition in anatase,

---
[a] Corresponding author at: Dept. of Materials Science and Engineering, Indian Institute of Technology (IIT) Gandhinagar, Palaj 382355, Gandhinagar, Gujarat, India.
 Electronic mail: emila@iitgn.ac.in

which would have significant impact in the photocatalytic and optoelectronic area.

## 1. Introduction

Anatase ($TiO_2$) is a chemically stable, low cost, non-toxic, wide band-gap semiconductor and has been mainly studied in the literature for the photocatalytic applications [1,2]. Presence of intrinsic defects could alter and/or create additional energy levels. In fact, depending on the position of these levels, either it could make this material even more attractive for photocatalytic application (by lowering the conduction band edge) or can make it an effective transparent conductor (TC) by adding charge carriers into the conduction and/or the valence bands without compromising on its transmittance. Moreover, it would be interesting to see if anatase can be designed where both these properties can co-exist. Out of all three thermodynamically stable phases, anatase is favoured both for the photocatalytic and the TC applications because of its relatively higher activity and lower electron effective mass as compared to others [3,4]. Though several experimental reports exist in the literature, most of these are found to focus only towards fabricating anatase for these applications, without providing a detailed scientific background [5,6] Fundamental understanding of this material system is necessary to design high quality materials for respective applications. Moreover, a large number of native point defects can be presented in anatase, which then modify its electronic structure, thereby significantly altering the optoelectronic properties. Furthermore, process parameters could significantly vary the microstructure of the synthesized materials, affecting their optoelectronic properties.

To this end, ab initio (first principle) calculations not only help in the understanding of how the defects affect the electronic structure of materials at the atomic level (and hence their optoelectronic properties), but also provide insights into how those properties might be improved. Though a range of first principle studies using density functional theory (DFT) have been carried out in the literature for the undoped and doped anatase, the obtained results were found to contradict, which could be attributed to the use of computational methodology in DFT [7]. In this regard, Phattalung *et al.* computed the native defects in anatase using Local Density

Approximation (LDA) exchange correlation functional and found none of these four defect states (i.e., oxygen vacancy ($V_O$), oxygen interstitial ($O_i$), titanium vacancy ($V_{Ti}$) and titanium interstitial ($Ti_i$)) being formed in the band gap [8]. However, experimental data had earlier revealed formation of mid-gap states in these systems, and therefore this discrepancy could be assigned to the incorrect self-interaction error of these functionals in DFT [9,10]. Note here that, the more advanced methods like DFT + U and hybrid functionals could correctly predict the existence of mid-gap defect levels due to native defects and could also predict the band gap accurately to a certain extent. To this end, Morgan *et al.* used GGA + U in their calculations and found localized mid-gap states being formed due to the neutral oxygen vacancies and titanium interstitials, both in anatase and in rutile [11]. J. Osorio Guillen *et al.* identified two types of behaviour for transition metal impurities in oxides; (i) in which the delocalized states were formed inside the conduction band thereby making the transparent material conductive and (ii) in which a localized mid gap state could form which then could transform the magnetic properties of the host material [12], indicating the energy of the outer *d* electrons of the impurity atom playing a decisive role on the eventual position of this defect state.

Most of the published works till date have investigated the effect of (neutral) native defects on the electronic structure of anatase. In this paper we used GGA + U approach in DFT to calculate the position and formation energies of various intrinsic (neutral as well as charged) defects for anatase. Apart from understanding the electronic structure of these systems individually, in this study formation energy of all these native defects was calculated to understand the stability of these defects under different conditions. Finally, an alignment of band diagrams for all the intrinsic defect states is performed using vacuum potential from slab-supercell calculation as reference.

## 2. Computational details

Here the electronic structure of pure anatase, along with those for the various intrinsic electronic defect states were calculated using restricted (spin-un-polarized) density functional theory (DFT) implemented in Quantum ESPRESSO software suite [13]. The neutral and charged states of the oxygen vacancies (denoted as $V_O$, $V_O^{+1}$, $V_O^{+2}$, thus the charge state varying from 0 to 2), oxygen interstitials ($O_i$), titanium vacancies ($V_{Ti}$) and titanium interstitials (denoted as $Ti_i$, $Ti_i^{+1}$, $Ti_i^{+2}$, $Ti_i^{+3}$, and $Ti_i^{+4}$, with the charge states varying from 0 to 4) were used as the intrinsic defect states. Here, a 2 × 2 × 2 supercell (constructed from the conventional cell of anatase) with a tetragonal structure (see Fig. 1(a)), which contained a total of 96 atoms (i.e., $Ti_{32}O_{64}$) was used for the computation. Moreover, here Generalized Gradient Approximation (GGA) Perdew-Burke-Ernzerhof (PBE) was adopted for the exchange-correlation potential [14]. Note that, standard DFT functionals tend to delocalize electrons over the crystal and hence, are not represented correctly only by these functionals, particularly for the material systems which contain transition elements with partially filled *d* or *f* orbitals. Hence along with GGA-PBE, a Hubbard-like (U) correction term, which accounted for the columbic repulsion of these localized *d* or *f* electrons was used here; henceforth denoted as DFT (GGA) + U approach [15]. The reported U value of 4.2 eV was used here to do the present calculation.

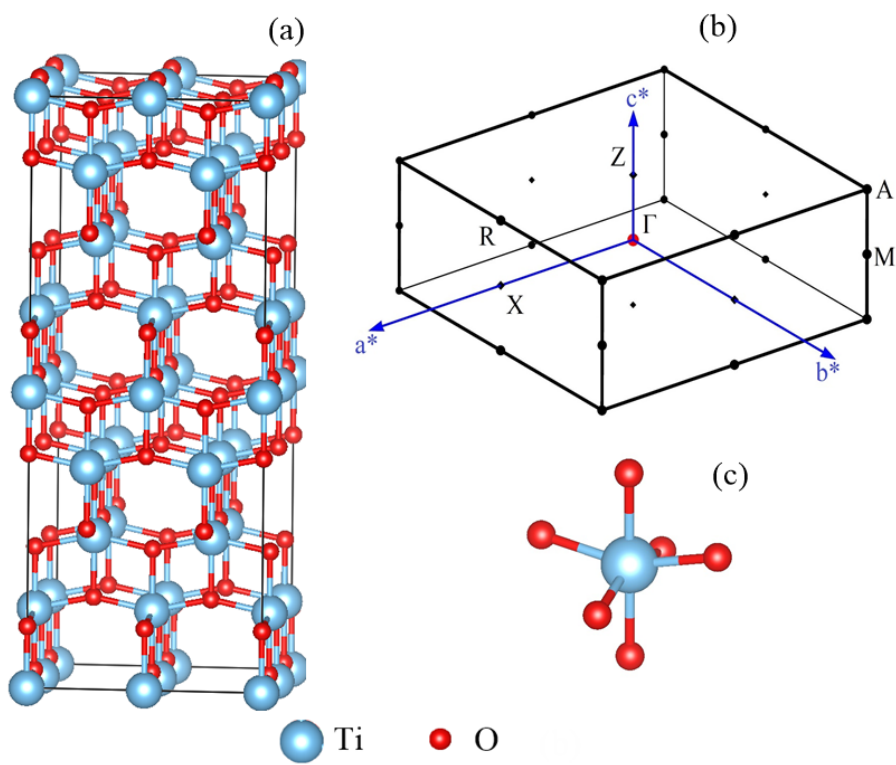

**Fig. 1.** Schematic of (a) anatase supercell, (b) the first brillouin zone of anatase lattice with high symmetry k points and (c) distorted $TiO_6$ octahedron of anatase. The two vertical Ti-O (apical) bonds are slightly longer than the other four equatorial Ti-O bonds.

Note that this value was used in the literature to calculate the lattice parameter, bulk modulus and defect states of surface oxygen vacancies in anatase using DFT (GGA) + U approach, and these calculated data were found to be in agreement with those of the experimental results and/or those calculated using the hybrid DFT route [16,17]. Here, projector augmented wave (PAW) type pseudopotentials [18] obtained from PSlibrary were used to model for the core electrons (i.e., [He], [Ne] for O and Ti atoms respectively) [19]. Further, $2s^2 2p^4$, and $3s^2 4s^2 3p^6 3d^2$ were used for O and Ti respectively in this computation as their valence electron configuration. Convergence tests were performed to determine the K-Points mesh for Brillouin zone integration and cut off energy for the plane wave expansion. It was found that moving from a Monkhorst-Pack grid of $6 \times 6 \times 6$ to a grid of $7 \times 7 \times 7$ changed the energy of the conventional anatase unit cell only by 0.0001 Ry. Hence, for a $2 \times 2 \times 2$ supercell, a $3 \times 3 \times 3$ mesh was considered as a reasonable value. A cut off energy of 55 Ry (i.e., 748 eV) was used for the plane wave expansion. Note that, increasing the cut off energy from 55 to 60 Ry changed the total energy of the cell by only about 0.001 Ry (i.e., 0.013 eV), thereby claiming 55 Ry as a reasonable cut off energy value. Moreover all the supercell structures were fully relaxed before proceeding with the calculations. The convergence threshold on energy and forces for ionic minimization were taken as $2.72 \times 10^{-3}$ eV and $5.14 \times 10^{-2}$ eV/Å respectively. Note that after performing the main calculation, a denser mesh of $6 \times 6 \times 6$ was used to compute the density of states (DOS) in order to obtain sharp peaks in the plot. For comparison of band gap of anatase, calculations using Tran and Blaha's modified Becke-Johnson (TB-mBJ) exchange potential [20,21] were also performed for pure anatase using Vienna *Ab initio* Simulation Package (VASP) [22–25] . Two calculations, were performed in this regard, one using mBJ exchange potential plus LDA correlation potential (mBJ-LDA) and other using mBJ exchange potential plus GGA-PBE correlation potential (mBJ-GGA). In both cases, relaxed anatase structure from GGA+U calculations was used and the pseudopotentials used were PAW type [26]. A cut off energy of 500 eV and a k-point mesh with a spacing of 0.2 Å$^{-1}$ was used.

Whereas intrinsic neutral defects of vacancies and interstitials were created by removing and adding one atom respectively from and to the supercell, charged defect states were created by using the charged supercell and a neutral background charge. For example, to create a neutral O vacancy ($V_O$), one neutral O atom was removed from the supercell, whereas, for creating $V_O^{+1}$, a neutral O atom and one electron were removed from the supercell and so on for other charged defects. To compare and understand the relative stabilities of various intrinsic and extrinsic defects, formation energy ($E^f[X^q]$) for all these above cases was calculated using the following equation:

$$E^f[X^q] = E_{tot}[X^q] - E_{tot}[bulk] - \sum_i n_i \mu_i + qE_F \qquad (1)$$

where, $E_{tot}[X^q]$ is the total energy of the supercell with defect $X^q$, $E_{tot}[bulk]$ is the total energy of an equivalent defect free supercell, $n_i$ represents the number of atoms of type $i$ that have been added to or removed from the supercell to create the defect, $\mu_i$ represents the chemical potential of the defect forming species (Ti or O), $q$ represents the charged state of the defect and $E_F$ is the Fermi level referenced to the valence band maxima (VBM) in the bulk.

To calculate the chemical potential of Ti and O in Ti- or O-rich conditions, the following boundary criteria were considered:

$$\mu_{Ti} + 2\mu_O = \mu_{TiO_2,bulk} \qquad (2)$$

$$\mu_{Ti} \leq \mu_{Ti,bulk} \qquad (3)$$

$$\mu_O \leq \frac{\mu_{O_2\ molecule}}{2} \qquad (4)$$

where, $\mu_{Ti}$ and $\mu_O$ are the chemical potentials of the defect forming species of Ti and O respectively; $\mu_{TiO_2,bulk}$, $\mu_{Ti,bulk}$ and $\mu_{O_2\ molecule}$ represent the chemical potentials of bulk $TiO_2$, Ti, and $O_2$ molecule respectively.

Note that, here the total energies were calculated for each of these structures per formula unit, which were then used as their chemical potential values. Moreover, $\mu_{Ti}$ and $\mu_O$ should always be lower than their natural phases of $\mu_{Ti,bulk}$ and $\mu_{O_2\ molecule}$ respectively, otherwise which these natural phases of Ti and $O_2$ would form instead of $TiO_2$ (see eq. (3) and (4)). Additionally, for O-rich (or Ti-poor) condition, $\mu_O = \frac{\mu_{O_2\ molecule}}{2}$ and for O-poor (or Ti-rich) condition, $\mu_{Ti} = \mu_{Ti,bulk}$.

The high symmetry points of the first brillouin zone which were used to plot the band structure (i.e., Γ at the centre of the brillouin zone, A at a vertex, R and M at the edge centres, X and Z at the face centres) are shown in Fig. 1(b). Next, the effective masses ($m^*$) of the electron and the hole were calculated by respectively fitting (parabolically) the bottom of the conduction band and top of the valence band along $\Gamma - X$ and $\Gamma - Z$ directions of the brillouin zone with the following relation:

$$m^* = \hbar \left(\frac{d^2E}{dk^2}\right)^{-1} \qquad (5)$$

where, $E$ is the energy of the electron at wave vector $k$ in a particular band, $\frac{d^2E}{dk^2}$ represents the curvature of that band and $\hbar$ is the reduced Planck's constant ($1.054 \times 10^{-34}$ J.s).

Finally, for aligning the band diagrams for different systems, vacuum level calculations were performed using the slab-supercell model in each case. Slabs with nine layers of atoms (with the atomic layers of (001) miller plane) and vacuum of 20 Å width on both the sides were used here. Convergence for the work function of pure anatase was checked with respect to the length of the vacuum and was found to be same up to the second decimal places irrespective of vacuum lengths of 19 and 20 Å, thereby making 20 Å a reasonable choice for vacuum level calculations. The work function of pure anatase was found to be 5.80 eV which is close to the experimentally measured vaue of 5.35 eV, indicating these calculations to have sufficient accuracy [27]. Since all the systems should have the same vacuum energy level, the vacuum energies were subtracted from the respective band energies for alignment.

## 3. Results and discussion

### 3.1. Electronic structure of pure anatase

The lattice parameters, Ti-O bond lengths of the distorted $TiO_6$ octahedron (see Fig. 1(c)), band gap and bader charges computed in this work are provided in Table 1, which were found to be in agreement with the literature reports [28–33]. The lattice parameters *a* and *b,* as well as the Ti-O bond lengths were in good agreement with literature, however there was a small deviation in the case of the lattice parameter *c*. In general, the computed lattice parameters depend upon the exchange-correlation functional and the pseudopotential used. Labat *et al.* investigated this in detail and concluded that for anatase, the error in lattice parameter *a* is ± 1 %, while *c* is systematically overestimated (up to 4 %) for all methods considered [34]. This is consistent with the values computed in the current work. This overall close agreement of the present data with those of the existed theoretical and experimental reports suggested reliability of the present computational work to predict the electronic structure and defect states of this system. The band structure (Fig. 2(a)) showed an indirect band gap ($E_g$) of 2.44 eV, which is an underestimation when compared to the experimental value. Whereas, conduction band in anatase consisted of mainly Ti 3d with a minor component of O 2p states, valence band was found to be mainly composed of O 2p with a minor contribution from Ti 3d states (see Fig. 2(b)). Note that, the mismatch in band gap arises primarily from the self interaction error in DFT, which over-delocalizes the partially occupied states and hence, forces them up in energy which in turn reduces the band gap. This effect is more pronounced in the case of highly localized $d$ and $f$ orbital electrons. GGA + U predicted a much better value of band gap than GGA alone (see Table 1), but it didn't correct the value completely. mBJ-GGA and mBJ-LDA both improved the band gap further to 2.6 eV and 2.8 eV respectively, with the latter giving the best prediction for the band gap of anatase. Although, mBJ-LDA gave the best value, all the defect calculations were performed using GGA + U because the relative positions of defect states and relative changes in band gaps were of more interest here, than their absolute values. As described in section 2, GGA + U (U = 4.2 eV) had previously found to correctly predict surface oxygen vacancy defect states, hence GGA + U was considered sufficient. There is also a mismatch in the band gap computed in the current work and other computational studies. This mismatch is probably due to the different methods and pseudopotentials used. Note that, the first study [33], utilised full potential (FP) linearized (L) augmented plane wave plus local orbitals (APW+lo) approach and GGA functional, the second [30] and the third studies [28] used GGA + U, but with different pseudopotentials: ultrasoft type and PAW type respectively.

The carrier effective mass of the electrons at the conduction band edge along $\Gamma - X$ and $\Gamma - Z$ direction was found to be 0.59 and 5.10 respectively. Similarly, for holes the effective mass values were 1.72 and 1.07 along the same directions. There was a large anisotropy in the effective mass (and hence the mobility) of electrons. Electrons were found to have an almost 10 times higher effective mass in $\Gamma - Z$ direction as compared to $\Gamma - X$ direction.

**Table 1**

Lattice parameters, bond lengths (in Å), band gap (in eV) and Bader charges for anatase

|  | Methodology | Lattice parameter | | | Ti-O bond length | | Band gap | Bader Charges | |
| --- | --- | --- | --- | --- | --- | --- | --- | --- | --- |
|  |  | $a$ | $b$ | $c$ | Apical | Equatorial |  | Ti | O |
| Current work | GGA + U | 3.84 | 3.84 | 9.84 | 1.99 | 1.95 | 2.44 | +2.25 | -1.15 |
|  |  |  |  |  |  |  | 2.6 (mBJ-GGA) |  |  |
|  |  |  |  |  |  |  | 2.8 (mBJ-LDA) |  |  |
| Other computational works | GGA [33] | 3.81 | 3.81 | 9.63 | _ | _ | 2.14 | _ | _ |
|  | GGA + U [30] | 3.83 | 3.83 | 9.63 | 2.00 | 1.96 | 2.00 | _ | _ |
|  | GGA + U [28] | 3.82 | 3.82 | 9.55 | _ | 1.95 | 2.61 | _ | _ |
|  | Full-potential all-electron calculations using PBE [29] | 3.81 | 3.81 | 9.72 | 2.01 | 1.95 | _ | +2.50 | -1.30 |
| Experimental value [31,32] |  | 3.78 | 3.78 | 9.50 | 1.98 | 1.93 | 3.21 | _ | _ |

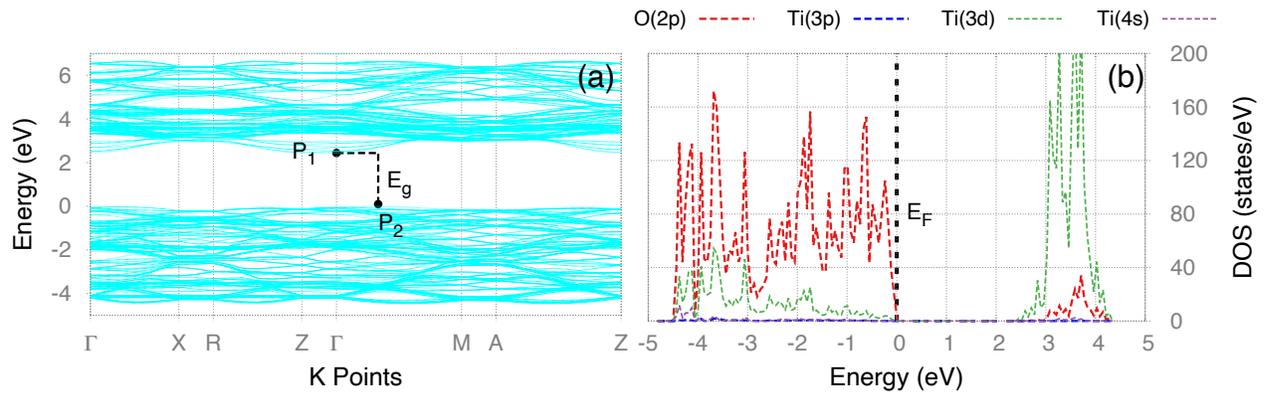

**Fig. 2.** Band structure (a) and PDOS (b) of anatase. Points P1 and P2 correspond to the CBM and VBM respectively and $E_g$ is the indirect band gap (2.44 eV). $E_F$ represents the fermi level. (For interpretation of the references to color in this figure legend, the reader is referred to the web version of this article.)

## 3.2. Native defect states in anatase

### 3.2.1. Geometrical distortion and charge redistribution due to neutral native defects

Here, the original bond lengths of the anatase crystal (i.e., of the undistorted lattice) were found to alter because of the creation of (neutral) native defects of $V_O$, $Ti_i$, $V_{Ti}$ and $O_i$, thereby bringing geometrical distortion to the vicinity of the defect sites (see Table 2 and Fig. 3). The directions in which the attractive and the repulsive forces were found to act on the neighbouring atoms of the native defects are shown in Fig. 3 and thus are known as defect-associated atoms.

**Table 2**

Distances (in Å) of the neighbouring atoms (atomic positions are illustrated in Fig. 3) from the defect site

| | Atomic distances from defect site X | | | | | | | | | | | |
|---|---|---|---|---|---|---|---|---|---|---|---|---|
| | $X = V_O$ | | | $X = Ti_i$ | | | $X = V_{Ti}$ | | | $X = O_i$ | | |
| | X-$Ti_{1,2}$ | X-$Ti_3$ | X-O | X-$O_{1,4}$ | X-$O_{2,3,5,6}$ | X-$Ti_{1-4}$ | X-$O_{1,4}$ | X-$O_{2,3,5,6}$ | X-Ti | X-$Ti_1$ | X-$Ti_2$ | X-$Ti_3$ |
| Pure anatase (undistorted lattice) | 1.95 | 1.99 | 2.48 | 1.84 | 2.25 | 2.42 | 1.99 | 1.95 | 3.06 | 1.95 | 1.99 | 1.95 |
| Anatase with X (distorted lattice) | 1.97 | 2.10 | 2.41 | 2.02 | 2.10 | 2.65 | 2.47 | 2.02 | 2.95 | 1.97 | 2.05 | 2.12 |

Three broken Ti-O bonds (i.e., $Ti_1$, $Ti_2$ and $Ti_3$ in Fig. 3(a)) were formed due to the creation of one $V_O$. Further, this $V_O$ in anatase was found to push these three nearest Ti atoms outwards because of their mutual strong repulsion, which could be associated to the higher oxidation state of Ti (see Table 2). The two nearby O atoms (labelled in Fig. 3(a)) were found to move towards $V_O$ in order to minimize the total energy of the system.

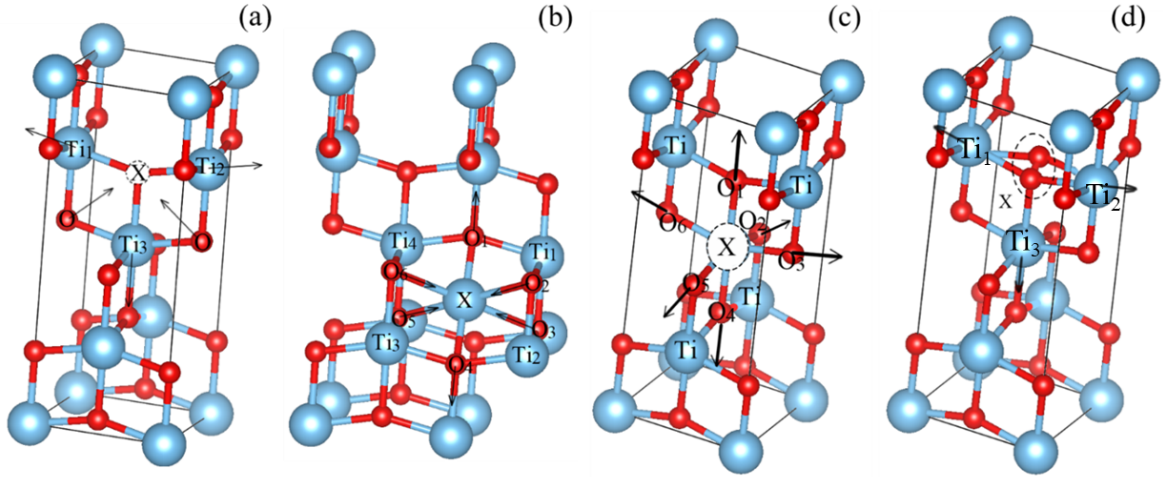

**Fig. 3.** A conventional cell of anatase showing various neutral native defects (denoted by X) such as $V_O$ (a), $Ti_i$ (b), $V_{Ti}$ (c) and $O_i$ (d). The approximate direction of forces acting on the neighbouring atoms is also depicted. The atoms in the neighbourhood (defect associated atoms) have been labelled.

Usually in literature, $Ti_i$ was seen to be placed at the centre of the distorted octahedron of the oxygen atoms in the anatase lattice [11]. However, we used two approaches in our calculation. In the first approach, $Ti_i$ was placed at a random location inside the crystal (position X in Fig. 3(b)) and allowed to relax to reach its lowest energy configuration. The second one followed the widely adopted literature approach, where the total energy calculation for anatase with $Ti_i$ at the centre of the distorted octahedron was carried out. The total energy of the first configuration was found to be slightly lower (by 0.23 eV) than that of the second one, suggesting increased stability of the first configuration over the second one. Hence, all further calculations here were carried out using the first configuration only. The neighbouring Ti atoms near $Ti_i$ ($Ti_1$, $Ti_2$, $Ti_3$ and $Ti_4$) were found to move outwards due to mutual repulsion among Ti atoms (see Table 2). Whereas, the two O atoms present in the apical positions (i.e., $O_1$ and $O_4$ in Fig. 3(b)) were seen to be pushed away from the interstitial site, the other four oxygen atoms in the neighbourhood (i.e., $O_2$, $O_3$, $O_5$ and $O_6$) were found to move towards this interstitial site due to the attraction from $Ti_i$ (see Table 2).

Moreover, removing a neutral Ti atom from anatase to create $V_{Ti}$ resulted in the six oxygen atoms in the neighbourhood (i.e., $O_1$-$O_6$) to relax outwards due to their strong mutual repulsion (see Fig. 3(c) and Table 2). Here, the oxygen atoms present at the apical positions (i.e., $O_1$ and $O_4$) were found to relax outwards by the maximum distance (~ 0.5 Å away from $V_{Ti}$; see Table 2) than those of the others. Ti atoms in this neighbourhood were found to be displaced inwards only slightly probably because of the missing repulsive force from the removed Ti atom.

Introducing $O_i$ in anatase resulted in the formation of a dimer with that of the lattice oxygen atom (see Fig. 3(d)). Note that, relaxation of atomic positions automatically resulted into this configuration even when we did not assume $O_i$ to form a dimer with a lattice oxygen atom. The O-O bond length of the dimer was found to be 1.47 Å, which is close to the O-O bond length in $[O_2]^{2-}$ as calculated for $BaO_2$ (1.49 Å), indicating the existence of the dimer in the form of $[O_2]^{2-}$.[35] The geometry in its neighbourhood was only found to be slightly affected, with the neighbouring Ti atoms ($Ti_1$, $Ti_2$, $Ti_3$) being moved slightly outwards (see Table 2).

**Table 3**

Partial charges on atoms at various positions (illustrated in Fig. 3) near defect X obtained from Bader charge analysis.

| Atoms | | Partial charge | | | |
|---|---|---|---|---|---|
| | | System with X = $V_O$ | System with X = $Ti_i$ | System with X = $V_{Ti}$ | System with X = $O_i$ |
| Defect atom X | | - | +1.83 | - | -0.58 |
| Ti atoms in the neighborhood of the defect X (defect associated atoms) | $Ti_1$ | +2.01 | +2.19 | - | - |
| | $Ti_2$ | +2.01 | +2.19 | - | - |
| | $Ti_3$ | +1.98 | +2.19 | - | - |
| | $Ti_4$ | - | +2.19 | - | - |
| O atoms in the neighbourhood of defect X (defect associated atoms) | $O_{1-6}$ | - | - | -0.93 | - |
| Ti atoms far away from defect X (average partial charge) | | +2.28 | +2.27 | +2.25 | +2.30 |
| O atoms far away from defect X (average partial charge) | | -1.15 | -1.17 | -1.14 | -1.15 |

Bader charge analysis was performed to know the distribution of charge (expressed in terms of the modulus of electronic charge $e$) on the neighbourhood atoms of native defects (labelled for all cases in Figs. 3(a)-3(d); see Table 3). In anatase with $V_O$, this analysis showed a reduction in the positive charge on defect associated Ti atoms ($Ti_{1-3}$) from +2.28 to ~ +2.01, indicating the unpaired electrons of the dangling bonds being localized over these atoms, adjacent to the vacant site (see Table 3). Charge density plot for the defect state clearly showed the extra charge being localized on the three nearby Ti atoms on Ti 3d orbitals for anatase with $V_O$ (see Fig. 4(a)). A significant amount of charge density was also observed in the position of $V_O$. Using GGA PW-91 functional in DFT for anatase with $V_O$, Kamisaka *et al*. had also found the electron density of the defect state to comprise of Ti 3d orbitals for the surrounding three Ti atoms [36].

Similarly, for anatase with $Ti_i$, the partial positive charge on the interstitial atom and that of the nearby Ti atoms ($Ti_{1-4}$) were found to be +1.83 and +2.19 respectively, as compared to that of +2.27 for Ti atoms, which were far away from the defect site, indicating the excess charge (due to the unpaired electrons) being primarily localized on the interstitial Ti atom and on the four Ti atoms in the neighbourhood of the interstitial site on the four-lobed Ti 3d orbitals (see Table 3 and Fig. 4(b)).

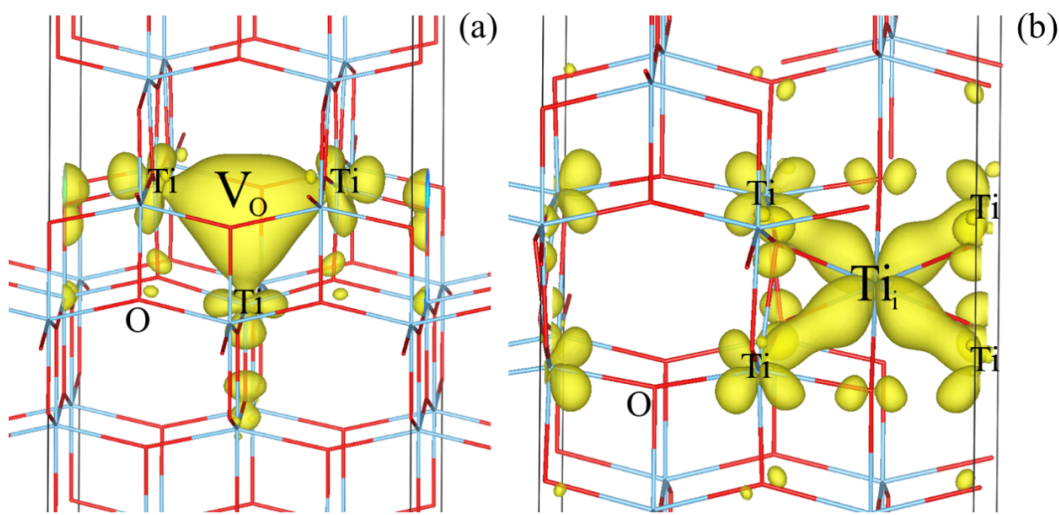

**Fig. 4.** Charge density plots (at Γ point) for anatase with $V_O$ (a), $Ti_i$ (b). The charge density has been plotted for the mid gap defect state. The isosurface around Ti atoms near $V_O$ and $Ti_i$ are shaped like $3d_{z^2}$ and $3d_{xy}$ orbitals respectively. The charge isosurface is shown at a density of 0.01 e/Å$^3$.

For anatase with $V_{Ti}$, the partial charge on the six O atoms ($O_{1-6}$) in the neighbourhood of the vacancy was found to be -0.93 as compared to the charge of -1.14 on O atoms away from the vacant site. This is expected because in a normal Ti-O bond, oxygen atom (because of its high electronegativity compared to Ti) would attract the bonding electrons towards itself which would create a partial negative charge on O atoms. So, when $V_{Ti}$ was created, the bond was broken, which decreased the partial negative charge on O atoms near the vacancy. Thus, these six O atoms were the primary defect associated atoms. In case of anatase with $O_i$, it was found that the partial charge on O atoms forming the dimer was ~ -0.58 as compared to the partial charge of -1.15 on O atoms, that were far from the defect site. O atoms in the dimer together had a partial charge of -1.16 which was found to be closer to the charge of -1.15 on O atoms away from the defect site, indicating charge similarity between $O_i$ defect site in its dimer state with that of a normal O lattice site, the only difference being the two oxygen atoms in $O_i$ instead of one.

### 3.2.2. Neutral and charged oxygen vacancies ($V_O$, $V_O^{+1}$, $V_O^{+2}$)

As oxygen atom has the valence state of two, $V_O$ in anatase would create two unpaired electrons in its vicinity. The DOS plots of anatase with various charged states of O vacancy clearly showed the formation of a mid-gap defect state (see Figs. 5(a)-5(d)). Further, this defect state was found to spread over a very small energy range, indicating this to be highly localized. Moreover, DOS plot of anatase with $V_O$ clearly showed the defect state being associated to Ti 3d orbitals, suggesting the states being localized on a few Ti atoms (see Fig. 5(b)). In this case, $E_F$ was found to lie at the edge of the defect state on the side of the conduction band, indicating this mid-gap states being occupied.

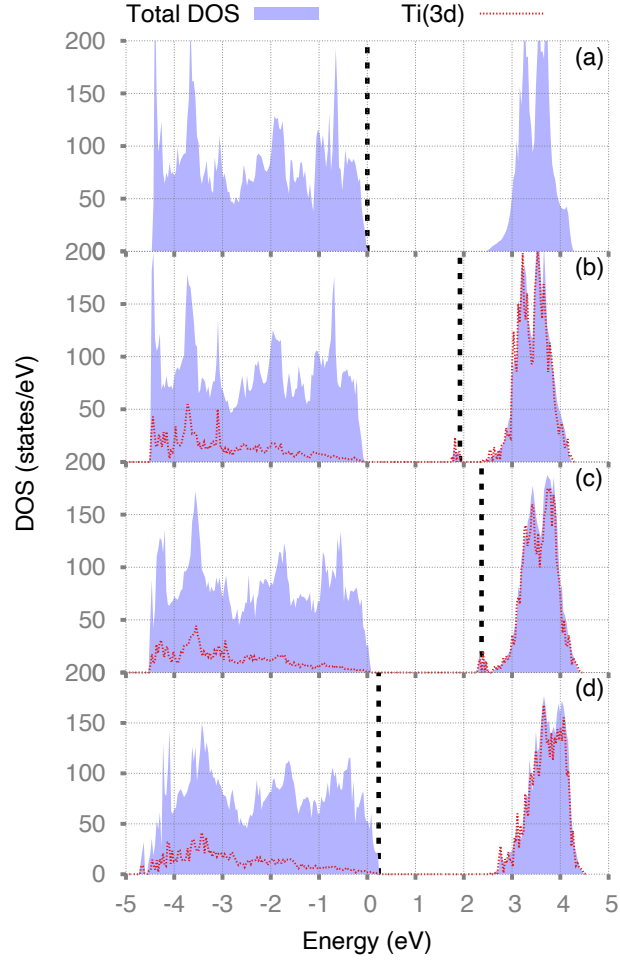

**Fig. 5.** DOS of pure anatase (a) and anatase with $V_O, V_O^{+1}, V_O^{+2}$ (b) – (d) respectively with Ti 3d states also shown. The mid gap defect state can be seen shifting towards the CBM as the charge on the vacancy increases. Black dotted line represents the $E_F$ and VBM of pure anatase has been chosen as the reference for all the plots. (For interpretation of the references to color in this figure legend, the reader is referred to the web version of this article.)

Whereas, for $V_O$ the defect states were found to form ~ 0.56 eV below the conduction band edge, the gap states were found to move closer to the conduction band edge with increasing charge on the vacancy from 0 to +2 (see Figs. 5(b)-5(d)). The gap state became very close to the conduction band edge (~ 0.07 eV) for anatase with $V_O^{+1}$. Additionally, $E_F$ in this case was found to be around the middle of the defect state (see Fig. 5(c)), hence, increasing the excitation probability of the electrons in the gap state to the conduction band and thereby providing *n*-type conductivity to anatase. For anatase with $V_O^{+2}$, the defect state was found to form inside the conduction band near the CBE (seen as a distinct peak at CBE in Fig. 5(d)) with $E_F$ being located at the VBM. The band gap values for anatase with $V_O, V_O^{+1}, V_O^{+2}$ were found to be 2.55 eV, 2.48 eV and 2.35 eV respectively. Note here the reduced band gap of anatase with $V_O^{+2}$ as compared to pure anatase, because of the formation of these defect states at the conduction band edge (see Fig. 5(d)). The band diagram of anatase with $V_o$ (see Fig. 6(a)) also showed the localized mid-gap defect state formed near the conduction band.

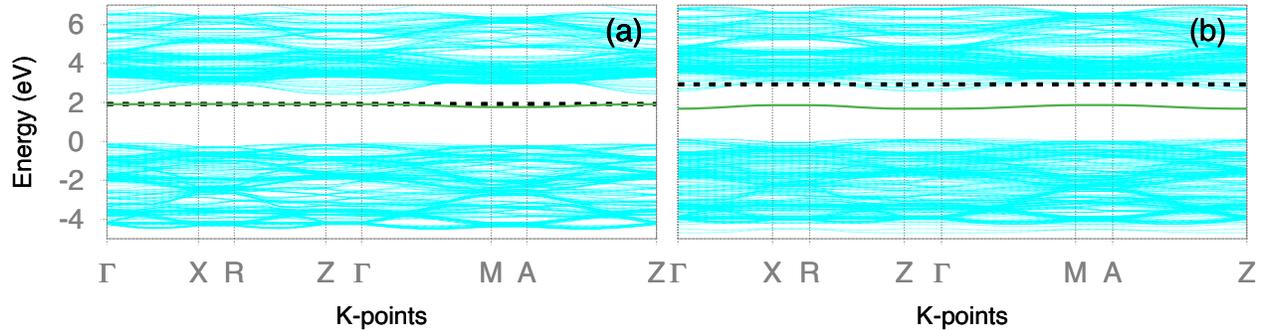

**Fig. 6.** Band diagram of anatase with (a) $V_O$ and (b) $Ti_i$ clearly showing the mid-gap defect states (green band). VBM of pure anatase has been taken as the reference energy value. Black dotted line shows Fermi level ($E_F$). (For interpretation of the references to color in this figure legend, the reader is referred to the web version of this article.)

### 3.2.3. Neutral and charged titanium interstitials ($Ti_i$, $Ti_i^{+1}$, $Ti_i^{+2}$, $Ti_i^{+3}$, $Ti_i^{+4}$)

A neutral Ti atom has a valence state of four, hence one $Ti_i$ in anatase would result into four unpaired electrons. The band diagram of anatase with $Ti_i$ clearly showed the formation of mid-gap defect state which was found to spread over a very small energy range indicating a higher degree of localization (see Fig. 6(b)). However, $E_F$ was found to lie in the conduction band, which is adjacent to CBE, indicating some of these excess charge carriers (or unpaired electrons of $Ti_i$) being delocalized, while the rest could be localized and present in the mid gap defect state.

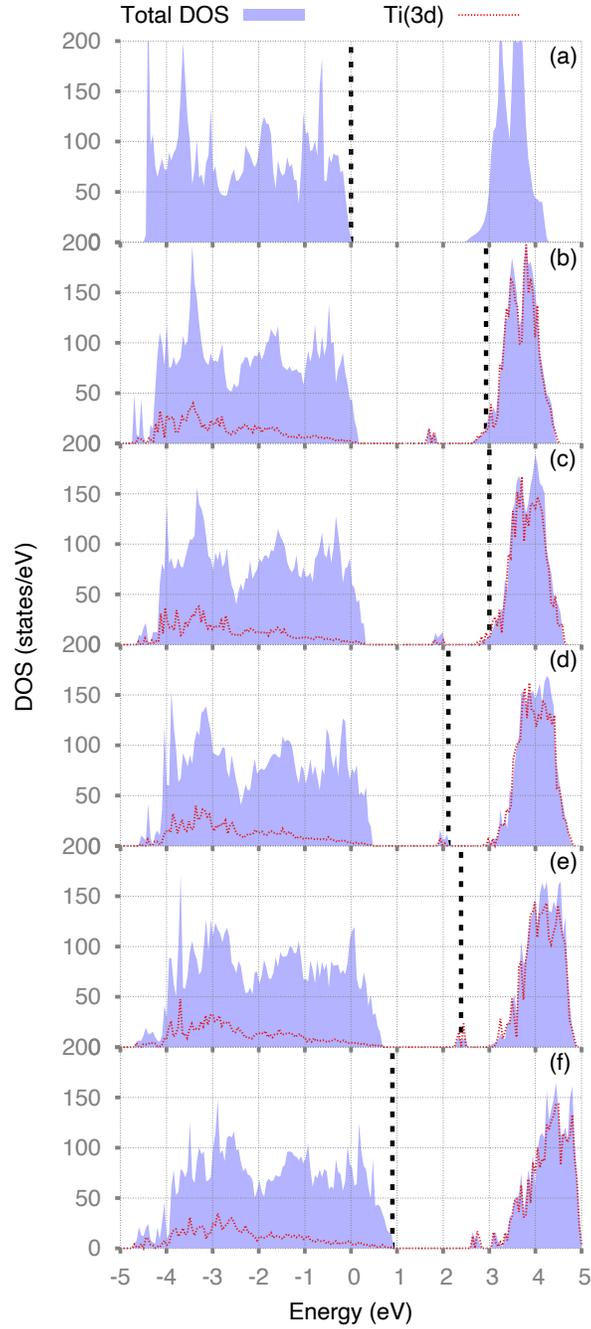

**Fig. 7.** (a) DOS of pure anatase and anatase with $Ti_i, Ti_i^{+1}, Ti_i^{+2}, Ti_i^{+3}$ and $Ti_i^{+4}$ (b) – (f) respectively with Ti 3d states also shown. Black dotted line represents the $E_F$. VBM of pure anatase has been taken as the reference energy value. (For interpretation of the references to color in this figure legend, the reader is referred to the web version of this article.)

DOS plots of anatase with different charge states of Ti interstitial clearly showed that this defect, in all its charged states created mid-gap defect states in the band gap which were associated

-17-

with Ti 3d orbitals (see Figs. 7(a)-7(f)). For $Ti_i^{+1}$, $E_F$ was still found to lie in the conduction band, however, the distance between it and the CBE decreased as compared to that of $Ti_i$ (i.e., 0.34 eV for $Ti_i$ and 0.21 eV for $Ti_i^{+1}$; see Figs. 7(b) and 7(c)). Further, in the case of $Ti_i^{+2}$, $E_F$ was found to locate at the edge of the mid gap state (see Fig. 7(d)), indicating the two electrons which were removed from $Ti_i$ to create $Ti_i^{+1}$ and $Ti_i^{+2}$ successively being present in the conduction band, hence suggesting the presence of the two defect states in the conduction band because of $Ti_i$. Moreover, when one more electron was removed to create $Ti_i^{+3}$, $E_F$ shifted to the middle of the mid gap defect state and finally for $Ti_i^{+4}$ it was found to coincide with the VBM (see Figs. 7(e)-7(f)). These observations indicate that the two electrons which were removed to create $Ti_i^{+3}$ and $Ti_i^{+4}$ successively from $Ti_i^{+2}$ occupied the localized mid gap defect states. Thus, $Ti_i$ was found to create two localized defect states in the band gap of anatase, whereas two delocalized states were formed in the conduction band, hence causing anatase with $Ti_i$ and/or $Ti_i^{+1}$, intrinsically *n*-type.

### 3.2.4. Neutral titanium vacancies ($V_{Ti}$) and oxygen interstitials ($O_i$)

In this calculation, no mid-gap defect states because of the presence of $V_{Ti}$ and/or $O_i$ in anatase were found to form (see Figs. 8(a)-8(d)). DOS plot for the defect associated O atoms showed O 2p states being formed inside the valence band and spread over a wide energy range indicating these as delocalized states (Figs. 8(c)-8(d)).

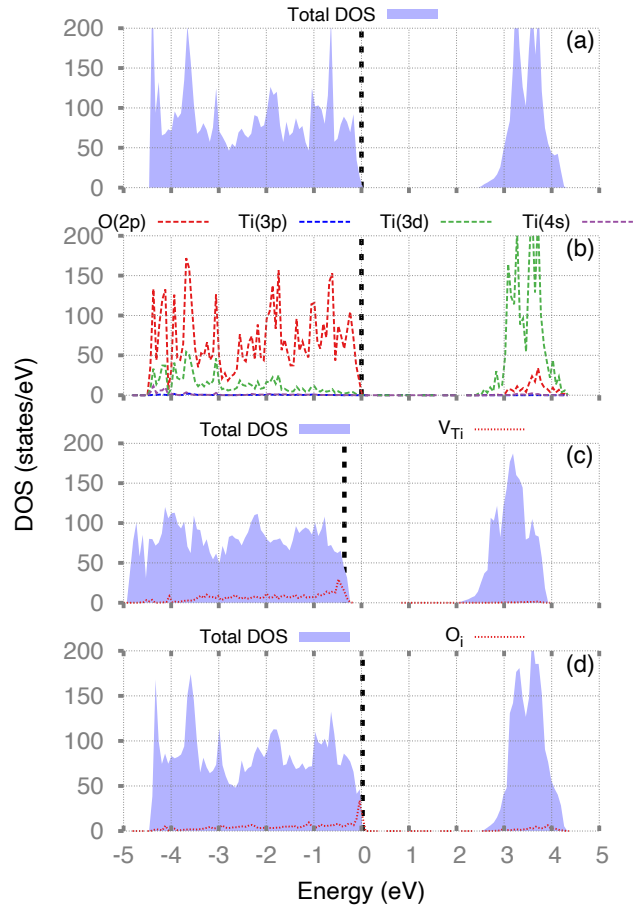

**Fig. 8.** Total DOS (a) and PDOS (b) of anatase. DOS of anatase with $V_{Ti}$ (c) and $O_i$ (d). Red dotted line in (c) and (d) represents O 2p states (scaled by 5 times) due to the defect associated atoms. Dotted line represents the $E_F$. VBM of pure anatase has been taken as the reference for all energies. (For interpretation of the references to color in this figure legend, the reader is referred to the web version of this article.)

*3.3. Stability of native defects*

The dominant defect types under various conditions were identified here by comparing their formation energies. Figs. 9(a) and 9(b) show the formation energies of native defects for O-poor and O-rich conditions respectively, obtained as a function of $E_F$, the upper and lower limits of which correspond to the CBM and VBM respectively. The charge transition levels (denoted by ε) of various native defects are shown in Table 4. Note that, these transition levels are $E_F$ values where multiple charged states of a defect state are stable.

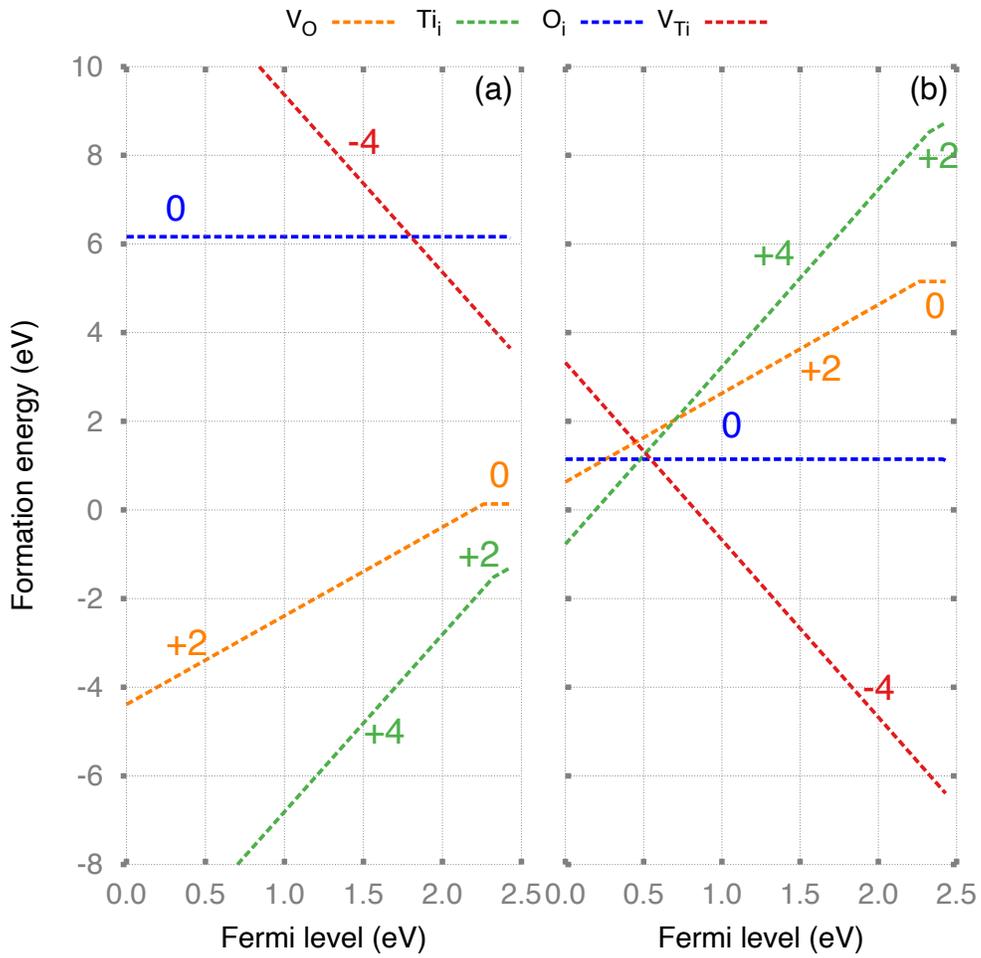

**Fig. 9.** Defect formation energies of native defects as a function of fermi level ($E_F$) in O-poor (a) and O-rich (b) conditions. The lower and upper limits of fermi level correspond to VBM and CBM respectively. (For interpretation of the references to color in this figure legend, the reader is referred to the web version of this article.)

**Table 4**

Charge transition levels (in eV) of various native defects with respect to the VBM.

| Defect | Charge states (q, q') | Transition level ε(q, q') |
|---|---|---|
| $V_O$ | +2/+1 | 2.27 |
| | +1/0 | 2.28 |
| $Ti_i$ | +4/+3 | 2.33 |
| | +3/+2 | 2.34 |
| | +2/+1 | 2.73 |
| | +1/0 | 2.74 |
| $V_{Ti}$ | 0/-1 | 0.01 |
| | -1/-2 | 0.02 |
| | -2/-3 | 0.03 |
| | -3/-4 | 0.04 |
| $O_i$ | 0/-1 | 2.42 |
| | -1/-2 | 2.43 |

For a wide range of $E_F$ (from 0 to ~2.2 eV), the most stable charged states of Ti vacancies, O vacancies, Ti interstitials and O interstitials were found to be -4, +2, +4 and 0 respectively (see Figs. 9(a) and 9(b)), suggesting Ti vacancies, O vacancies and Ti interstitials being acted as quadruple acceptor, double donor, and quadruple donor respectively. Both in O-poor and in O-rich conditions, the most stable defect states were found to be $Ti_i^{+4}$ and $V_{Ti}^{-4}$. Moreover, in O-poor condition (see Fig. 9(a)), $Ti_i^{+4}$ was found to be the most stable defect state for the entire range of $E_F$. In our calculation, O interstitials were found to have a positive formation energy for both these conditions and thus were unlikely to form spontaneously under equilibrium conditions. For a wide range of $E_F$ values (from 0 to ~2.4 eV), neutral state of O interstitials was found to be more stable as compared to its other charged states and hence were likely to bind with a lattice oxygen atom and form a dimer configuration. Although oxygen vacancies were not the most stable defect states in any of the two conditions, these in the form of $V_O^{+2}$ were found to have a negative formation energy for $E_F$ ranging between 0 to 2.2 eV in O-poor conditions (Fig. 9(a)), and hence are likely to form.

The charge transition levels of O vacancies ($\varepsilon(+2, +1)$ and $\varepsilon(+1, 0)$) were found to lie only slightly below the CBM (~ 0.16 eV; see Table 4) indicating these as shallow donor defects. As a result, these electrons could ionize easily and thereby leading to intrinsically *n*-type conductivity in anatase. Similarly, out of four transition levels of $Ti_i$, two (i.e., $\varepsilon(+4, +3)$ and $\varepsilon(+3, +2)$) were found to locate close to the CBM, whereas the other two (i.e., $\varepsilon(+2, +1)$ and $\varepsilon(+1, 0)$) lied in the conduction band, thereby clearly indicating Ti interstitial to provide *n*-type conductivity to anatase. Note that, the transition levels of Ti vacancies (from $\varepsilon(0, -1)$ to $\varepsilon(-3, -4)$) were found to lie almost at the valence band edge, indicating these as shallow acceptors. Finally, the transition levels of O interstitials ($\varepsilon(0, -1)$) and $\varepsilon(-1, -2)$ were found to lie ~ 0.02 eV below the CBM. Moreover the stability of -2 charged state of O interstitials near the CBM could imply O interstitials being present as the acceptor states, however note that these would be very unlikely to form due to their high formation energy.

Note that, in O-poor condition, donor type defects (O vacancy and Ti interstitial) were found to have a lower formation energy than those of the acceptor type defects (Ti vacancy and O interstitial) for the entire range of $E_F$ (see Fig. 9(a)). This could then lead to an incomplete compensation of the electrons (induced by the donor-type defects) by the holes (induced by the acceptor-type defects), thereby making anatase intrinsically *n*-type, reason behind the growth of intrinsically *n*-type anatase for oxygen deficient samples. However, in O-rich conditions, the acceptor-type defect states (mainly Ti vacancy) was found to be more stable, which could make anatase *p*-type (see Fig. 9(b)) under these conditions.

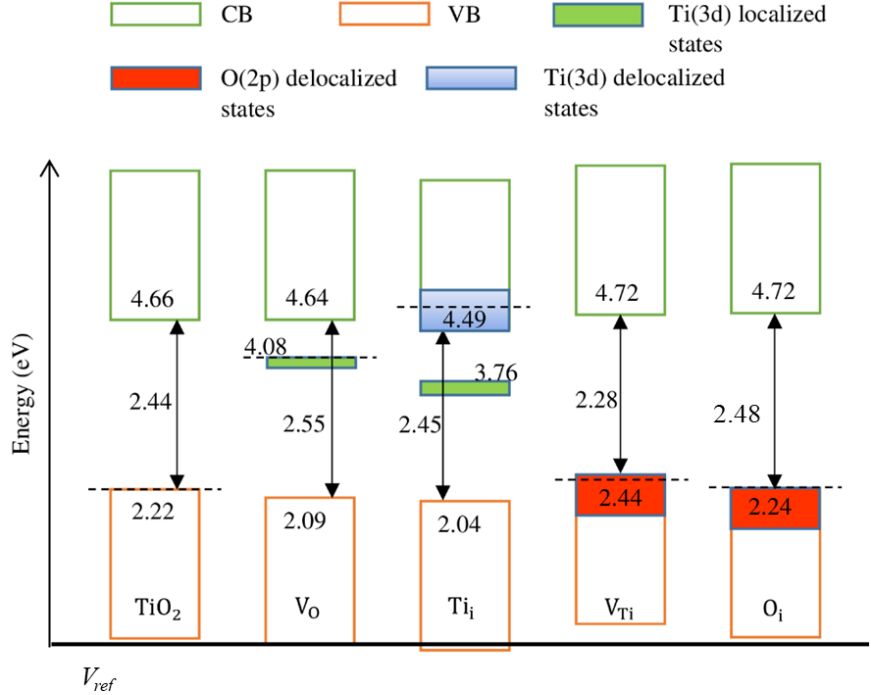

**Fig. 10.** Band line-up diagram for bulk systems of pure anatase, anatase with $V_O$, $Ti_i$, $V_{Ti}$ and $O_i$. The positions of band edges and the band gap value have been shown. Both the localized and delocalized defect states are shown. Dotted line represents $E_F$. Vacuum level calculated using the slab configuration was used as the reference ($V_{ref}$) for energies for different bulk systems. (For interpretation of the references to color in this figure legend, the reader is referred to the web version of this article.)

*3.4. Shifts in VBE and CBE due to neutral native defects*

Fig. 10 shows the band line-up diagram for various bulk systems using vacuum (calculated from slab configuration) as the reference potential ($V_{ref}$). The conduction band edge (CBE) and the valence band edge (VBE) of pure anatase were found to lie at 4.66 eV and 2.22 eV respectively. For anatase with $V_O$, VBE was found to shift downward by 0.13 eV, whereas the CBE shifted upward very slightly. The downward shift of the VBE could be attributed to the less band width of O 2p states because of the removal of one oxygen atom. For anatase with $Ti_i$, CBE was found to shift downwards as in addition to forming localized mid gap state, $Ti_i$ also formed delocalized defect states at the edge of the conduction band (see Sec. III B.3). For anatase with $V_{Ti}$ and also with $O_i$, VBE shifted upward because both these native defects formed O 2p delocalized states in the valence band. In these cases, the CBE was also found to move upward. The upward shift of CBM in $V_{Ti}$ could be because of less band width of Ti 3d states because of the removal of one Ti atom.

-23-

## 4. Conclusions

In this work an overall electronic structure including the position and formation energies of various intrinsic point defects were computed for anatase using DFT+U. $V_O$, $O_i$, $V_{Ti}$ and $Ti_i$ were considered here as the native defects. Additionally, in this study effective mass of the electrons and holes for anatase were calculated from the curvature of their *E-K* diagram.

Pure anatase band gap was found to be 2.44 eV (an underestimation). Whereas, the conduction band in undoped anatase was found to consist of mainly Ti 3d states with a minor component of O 2p states, the valence band was found to be mainly composed of O 2p states with a minor contribution from Ti 3d states. Anisotropy in the band structure was found to exist, which led to widely different values of effective masses of electrons moving along different directions in the brillouin zone.

Calculations for the native defects showed that O vacancies (for charged states 0, +1) and Ti interstitials (for all charged states) were associated with localized Ti 3d defect states in the band gap, whereas Ti vacancies and O interstitials formed delocalized O 2p states in the valence band. The gap state was quite close to the conduction band edge in case of $V_O^{+1}$ and along with localized states, Ti interstitials also formed delocalized states in the conduction band (with fermi level located inside the conduction band for $Ti_i$ and $Ti_i^{+1}$). These factors explained the intrinsic conductivity in anatase. The most stable charged states for Ti and O vacancies, Ti interstitials were found to be -4, +2 and +4 respectively, indicating these as quadruple acceptor, double donor and quadruple donor respectively. O interstitial was found to form a stable dimer configuration with a lattice O atom. Moreover, oxygen deficient anatase was predicted here to be intrinsically *n*-type, whereas *p*-type anatase was predicted in O-rich conditions. Finally, an alignment of band diagrams for all the intrinsic defect states was carried out using slab-supercell calculation and employing vacuum as the reference potential.

## Acknowledgements

The computations in this work have been performed using the facilities of Research Center for Advanced Computing Infrastructure at JAIST. Ryo Maezono is grateful for financial supports from MEXT-KAKENHI (17H05478 and 16KK0097), from FLAGSHIP2020 (RIKEN, project nos. hp190169 and hp190167 at K-computer), from I-O DATA Foundation, from the Air Force Office of Scientific Research (AFOSR-AOARD/FA2386-17-1-4049), and from JSPS Bilateral Joint Projects (with India DST). Emila Panda gratefully acknowledges financial support from Science and Engineering Research Board, Department of Science and Technology, Government of India (Project no: EMR/2016/001182). Abhishek Raghav would like to gratefully acknowledge the financial support from the Ministry of Human Resource and Development (MHRD), Government of India, for conducting this work at IIT Gandhinagar (IITGN) and Japan Student Services Organization (JASSO) for providing partial financial support for this project. The funding sources had no involvement in study design, in the writing of article, and in the decision to submit the article for preparation.
## Data availability

The raw data required to reproduce these findings are available within the article.

-25-

# References

[1]     A. FUJISHIMA, K. HONDA, Electrochemical Photolysis of Water at a Semiconductor Electrode, Nature. 238 (1972) 37–38. https://doi.org/10.1038/238037a0.

[2]     K. Nakata, A. Fujishima, TiO2 photocatalysis: Design and applications, J. Photochem. Photobiol. C Photochem. Rev. 13 (2012) 169–189. https://doi.org/https://doi.org/10.1016/j.jphotochemrev.2012.06.001.

[3]     T. Luttrell, S. Halpegamage, J. Tao, A. Kramer, E. Sutter, M. Batzill, Why is anatase a better photocatalyst than rutile? - Model studies on epitaxial TiO2 films, Sci. Rep. 4 (2015) 4043. https://doi.org/10.1038/srep04043.

[4]     J. Zhang, P. Zhou, J. Liu, J. Yu, New understanding of the difference of photocatalytic activity among anatase, rutile and brookite TiO 2, Phys. Chem. Chem. Phys. 16 (2014) 20382–20386. https://doi.org/10.1039/C4CP02201G.

[5]     N. Yamada, T. Hitosugi, N.L.H. Hoang, Y. Furubayashi, Y. Hirose, T. Shimada, T. Hasegawa, Fabrication of Low Resistivity Nb-doped TiO 2 Transparent Conductive Polycrystalline Films on Glass by Reactive Sputtering, Jpn. J. Appl. Phys. 46 (2007) 5275–5277. https://doi.org/10.1143/JJAP.46.5275.

[6]     C. Das, P. Roy, M. Yang, H. Jha, P. Schmuki, Nb doped TiO 2 nanotubes for enhanced photoelectrochemical water-splitting, Nanoscale. 3 (2011) 3094–3096.

[7]     E. Finazzi, C. Di Valentin, G. Pacchioni, A. Selloni, Excess electron states in reduced bulk anatase TiO2: Comparison of standard GGA, GGA+U, and hybrid DFT calculations, J. Chem. Phys. 129 (2008) 154113. https://doi.org/10.1063/1.2996362.

[8]     S. Na-Phattalung, M.F. Smith, K. Kim, M.H. Du, S.H. Wei, S.B. Zhang, S. Limpijumnong, First-principles study of native defects in anatase Ti O2, Phys. Rev. B - Condens. Matter Mater. Phys. 73 (2006) 125205. https://doi.org/10.1103/PhysRevB.73.125205.

[9]     N.D. Abazović, M.I. Čomor, M.D. Dramićanin, D.J. Jovanović, S.P. Ahrenkiel, J.M. Nedeljković, Photoluminescence of Anatase and Rutile TiO 2 Particles †, J. Phys. Chem.

B. 110 (2006) 25366–25370. https://doi.org/10.1021/jp064454f.

[10] T. Berger, M. Sterrer, O. Diwald, E. Knözinger, D. Panayotov, T.L. Thompson, J.T. Yates, Light-Induced Charge Separation in Anatase TiO 2 Particles, J. Phys. Chem. B. 109 (2005) 6061–6068. https://doi.org/10.1021/jp0404293.

[11] B.J. Morgan, G.W. Watson, Intrinsic n-type Defect Formation in TiO 2 : A Comparison of Rutile and Anatase from GGA+ U Calculations, J. Phys. Chem. C. 114 (2010) 2321–2328. https://doi.org/10.1021/jp9088047.

[12] J. Osorio-Guillén, S. Lany, A. Zunger, Atomic control of conductivity versus ferromagnetism in wide-gap oxides via selective doping: V, Nb, Ta in anatase TiO2, Phys. Rev. Lett. 100 (2008) 036601. https://doi.org/10.1103/PhysRevLett.100.036601.

[13] P. Giannozzi, S. Baroni, N. Bonini, M. Calandra, R. Car, C. Cavazzoni, D. Ceresoli, G.L. Chiarotti, M. Cococcioni, I. Dabo, A. Dal Corso, S. De Gironcoli, S. Fabris, G. Fratesi, R. Gebauer, U. Gerstmann, C. Gougoussis, A. Kokalj, M. Lazzeri, L. Martin-Samos, N. Marzari, F. Mauri, R. Mazzarello, S. Paolini, A. Pasquarello, L. Paulatto, C. Sbraccia, S. Scandolo, G. Sclauzero, A.P. Seitsonen, A. Smogunov, P. Umari, R.M. Wentzcovitch, QUANTUM ESPRESSO: A modular and open-source software project for quantum simulations of materials, J. Phys. Condens. Matter. 21 (2009) 395502. https://doi.org/10.1088/0953-8984/21/39/395502.

[14] J.P. Perdew, K. Burke, M. Ernzerhof, Erratum: Generalized gradient approximation made simple (Physical Review Letters (1996) 77 (3865)), Phys. Rev. Lett. 78 (1997) 1396. https://doi.org/10.1103/PhysRevLett.78.1396.

[15] C. Freysoldt, B. Grabowski, T. Hickel, J. Neugebauer, G. Kresse, A. Janotti, C.G. Van De Walle, First-principles calculations for point defects in solids, Rev. Mod. Phys. 86 (2014) 253–305. https://doi.org/10.1103/RevModPhys.86.253.

[16] S.K. Gharaei, M. Abbasnejad, R. Maezono, Bandgap reduction of photocatalytic TiO2 nanotube by Cu doping, Sci. Rep. 8 (2018). https://doi.org/10.1038/s41598-018-32130-w.

[17] B.J. Morgan, D.O. Scanlon, G.W. Watson, Small polarons in Nb- and Ta-doped rutile and anatase TiO2, J. Mater. Chem. 19 (2009) 5175–5178. https://doi.org/10.1039/b905028k.

[18]   P.E. Blöchl, Projector augmented-wave method, Phys. Rev. B. 50 (1994) 17953–17979. https://doi.org/10.1103/PhysRevB.50.17953.

[19]   A. Dal Corso, Pseudopotentials periodic table: From H to Pu, Comput. Mater. Sci. 95 (2014) 337–350. https://doi.org/10.1016/j.commatsci.2014.07.043.

[20]   A.D. Becke, E.R. Johnson, A simple effective potential for exchange, J. Chem. Phys. (2006). https://doi.org/10.1063/1.2213970.

[21]   F. Tran, P. Blaha, Accurate band gaps of semiconductors and insulators with a semilocal exchange-correlation potential, Phys. Rev. Lett. (2009). https://doi.org/10.1103/PhysRevLett.102.226401.

[22]   G. Kresse, J. Hafner, Ab initio molecular dynamics for liquid metals, Phys. Rev. B. (1993). https://doi.org/10.1103/PhysRevB.47.558.

[23]   G. Kresse, J. Hafner, Ab initio molecular-dynamics simulation of the liquid-metal–amorphous-semiconductor transition in germanium, Phys. Rev. B. 49 (1994) 14251–14269. https://doi.org/10.1103/PhysRevB.49.14251.

[24]   G. Kresse, J. Furthmüller, Efficiency of ab-initio total energy calculations for metals and semiconductors using a plane-wave basis set, Comput. Mater. Sci. 6 (1996) 15–50. https://doi.org/10.1016/0927-0256(96)00008-0.

[25]   G. Kresse, J. Furthmüller, Efficient iterative schemes for ab initio total-energy calculations using a plane-wave basis set, Phys. Rev. B. 54 (1996) 11169–11186. https://doi.org/10.1103/PhysRevB.54.11169.

[26]   G. Kresse, D. Joubert, From ultrasoft pseudopotentials to the projector augmented-wave method, Phys. Rev. B. 59 (1999) 1758–1775. https://doi.org/10.1103/PhysRevB.59.1758.

[27]   S. Kashiwaya, J. Morasch, V. Streibel, T. Toupance, W. Jaegermann, A. Klein, The Work Function of $TiO_2$, Surfaces. 1 (2018) 73–89. https://doi.org/10.3390/surfaces1010007.

[28]   E. German, R. Faccio, A.W. Mombrú, A DFT + U study on structural, electronic, vibrational and thermodynamic properties of $TiO_2$ polymorphs and hydrogen titanate: Tuning the hubbard 'U-term,' J. Phys. Commun. 1 (2017) 055006.

https://doi.org/10.1088/2399-6528/aa8573.

[29] D. Koch, S. Manzhos, On the Charge State of Titanium in Titanium Dioxide, J. Phys. Chem. Lett. 8 (2017) 1593–1598. https://doi.org/10.1021/acs.jpclett.7b00313.

[30] E. Araujo-Lopez, L.A. Varilla, N. Seriani, J.A. Montoya, TiO2 anatase's bulk and (001) surface, structural and electronic properties: A DFT study on the importance of Hubbard and van der Waals contributions, Surf. Sci. 653 (2016) 187–196. https://doi.org/10.1016/j.susc.2016.07.003.

[31] D. Reyes-Coronado, G. Rodríguez-Gattorno, M.E. Espinosa-Pesqueira, C. Cab, R. De Coss, G. Oskam, Phase-pure TiO2 nanoparticles: Anatase, brookite and rutile, Nanotechnology. 19 (2008) 145605. https://doi.org/10.1088/0957-4484/19/14/145605.

[32] J.K. Burdett, T. Hughbanks, G.J. Miller, J. V. Smith, J.W. Richardson, Structural-Electronic Relationships in Inorganic Solids: Powder Neutron Diffraction Studies of the Rutile and Anatase Polymorphs of Titanium Dioxide at 15 and 295 K, J. Am. Chem. Soc. 109 (1987) 3639–3646. https://doi.org/10.1021/ja00246a021.

[33] M. Mohamad, B.U. Haq, R. Ahmed, A. Shaari, N. Ali, R. Hussain, A density functional study of structural, electronic and optical properties of titanium dioxide: Characterization of rutile, anatase and brookite polymorphs, Mater. Sci. Semicond. Process. 31 (2015) 405–414. https://doi.org/10.1016/j.mssp.2014.12.027.

[34] F. Labat, P. Baranek, C. Domain, C. Minot, C. Adamo, Density functional theory analysis of the structural and electronic properties of TiO2 rutile and anatase polytypes: Performances of different exchange-correlation functionals, J. Chem. Phys. 126 (2007) 154703. https://doi.org/10.1063/1.2717168.

[35] H. Dietrich, Tables of Interatomic Distances and Configuration in Molecules and Ions, herausgeg. von A. D. Mitchell und L. C. Cross. Special Publication No. 11. Wiss. Herausgeber: L. E. Sutton. The Chemical Society, London 1958. 1. Aufl., 385 S., geb. £ 2.2.0, Angew. Chemie. 73 (1961) 511–512. https://doi.org/10.1002/ange.19610731425.

[36] H. Kamisaka, T. Hitosugi, T. Suenaga, T. Hasegawa, K. Yamashita, Density functional theory based first-principle calculation of Nb-doped anatase TiO2 and its interactions with

oxygen vacancies and interstitial oxygen, J. Chem. Phys. 131 (2009). https://doi.org/10.1063/1.3157283.

## DECLARATION OF INTERESTS

The authors declare that they have no known competing financial interests or personal relationships that could have appeared to influence the work reported in this paper.